# A new approach for improving global critical current density in Fe(Se$_{0.5}$Te$_{0.5}$) polycrystalline materials

A.Palenzona[1,3], A.Sala[1,2], C. Bernini[3], V. Braccini[3], M.R.Cimberle[4], C.Ferdeghini[3], G. Lamura[3], A.Martinelli[3], I. Pallecchi[3], G. Romano[3], M.Tropeano[1,3], R.Fittipaldi[5], A.Vecchione[5], A. Polyanskii[6], F. Kametani[6], M. Putti[2,3]

[1] Chemistry and Industrial Chemistry Department, University of Genova, via Dodecaneso 31, I-16146 Genova, Italy

[2] Physics Department, University of Genova, via Dodecaneso 33, I-16146 Genova, Italy

[3] CNR-SPIN-Genova corso Perrone 24, 16152, Genova, Italy

[4] CNR-IMEM Physics Department, University of Genova, via Dodecaneso 33, I-16146 Genova, Italy

[5] CNR-SPIN-Salerno and Physics Department University of Salerno, Via Ponte don Melillo, I-84084 Fisciano (SA), Italy

[6] Applied Superconductivity Center, National High Magnetic Field Laboratory 2031 E. Paul Dirac, Dr., Tallahassee, FL 32310, USA

E-mail: alberto.sala@spin.cnr.it

**Abstract.** A novel method to prepare bulk Fe(Se$_{0.5}$Te$_{0.5}$) samples is presented, based on a melting process and a subsequent annealing treatment. With respect to the standard sintering technique, it produces much more homogeneous and denser samples, characterized by large and well interconnected grains. The resulting samples exhibit optimal critical temperature values, sharp resistive and magnetic transitions, large magnetic hysteresis loops and high upper critical fields are observed. Interestingly, the global critical current density is much enhanced as compared to the values reported in literature for bulk samples of the same 11 family, reaching about $10^3$ A/cm$^2$ at zero field at 4.2 K as assessed by magnetic, transport and magneto-optical techniques. Even more importantly, its field dependence turns out to be very weak, such that at $\mu_0 H = 7$ T it is suppressed only by a factor ~2.

# 1. Introduction

The discovery of superconductivity in iron-based pnictides and chalcogenides has triggered the interest of the international community toward possible applicative issues: these materials result technologically very appealing for high field applications, being characterized by high critical superconducting temperature ($T_c$) and huge upper critical field ($B_{c2}$) values [1]. The critical current density $J_c$ of single crystal samples is particularly promising since it combines high magnitude and nearly isotropic character [2,3,4,5,6]. Instead, the global critical current density, $J_c^{global}$, measured in bulk samples has appeared rather low since the beginning [7] even if larger than in randomly oriented polycrystalline cuprates [8]. It strongly decreases by applying a magnetic field [9], which is a signature of Josephson coupling between the grains. The weak links between the grains are mainly related to the microstructure of the samples. Microstructural investigations have emphasized cracks, grain boundary-wetting Fe–As phase, [9] and phase inhomogeneities which cause local suppression of the order parameter at grain boundaries [10] as the main mechanisms responsible for current blocking in polycrystalline materials.

In addition, measurements on thin films grown on bicrystals [11,12] have indicated that $J_c$ exponentially decreases with the misorientation angle, with a behaviour reminiscent of that already observed in high-$T_c$ cuprates [13], thus suggesting intrinsic limitations to achieve high $J_c^{global}$ values in bulk materials. However, $J_c$ across grain boundaries in actual conductors is often better than in grain boundaries of epitaxial films grown on bicrystals, as observed in a recent review on this subject [14]. This is also confirmed by results recently obtained mainly on the 122 phase: in particular, a critical current density of the order of $10^4$ A/cm$^2$ in self-field and $10^3$ A/cm$^2$ at 10 T has been obtained in Sr$_{1-x}$K$_x$Fe$_2$As$_2$ textured tapes fabricated by a cold deformation process plus addition of a metallic element, which are both effective ways to improve grain connectivity [15]. Similar results have been obtained by Ding et al. on Ba$_{0.6}$K$_{0.4}$Fe$_2$As$_2$ wires performed by the ex-situ Powder in Tube (PIT) technique[16]. Even larger values ($10^5$ A/cm$^2$ in self-field and $10^4$ A/cm$^2$ at 10 T) have been reported by J. D. Weiss et al. in fine grain Ba$_{0.6}$K$_{0.4}$Fe$_2$As$_2$ wires and bulk [17].

In the 11 phase, on the other hand, attempts to improve the global $J_c$ have been less successful. Fe(Se$_{0.5}$Te$_{0.5}$) wires have been synthesized by the Powder In Tube (PIT) technique obtaining a transport $J_c$ value of $10^2$ A/cm$^2$ [18] in the favourable hypothesis of current flowing in the external edge only. In these samples $J_c$ is suppressed by one order of magnitude at $\mu_0 H = 1$T. Ozaki et al. [19] have measured $J_c$ values of the same order of magnitude at T = 4.2 K in self-field in Fe(Se$_{0.5}$Te$_{0.5}$) wires produced by a simple PIT technique. A better result has been obtained by the same authors in FeSe wires [20], where $J_c$ reaches $10^3$ A/cm$^2$ at T = 4.2 K in self field, but becomes two orders of magnitude lower at 10 T. Ding et al. in FeSe tapes prepared by using a diffusion method achieved a transport critical current density as high as 600 A cm$^2$ at 4.2 K under self-field, and observed a distribution of Tc and weak-link features by magneto-optical imaging [21]. The same group has observed 700 A/cm$^2$ in self field and T = 4.2 K in polycrystalline samples prepared by solid state reaction method, but a very severe drop with the field occurs, namely $J_c$ decreases by more than an order of magnitude under an applied of 0.5 T only [22].

Here we present an original melting process to prepare Fe(Se$_{0.5}$Te$_{0.5}$) bulk material By this process it is possible to fabricate much denser and more homogeneous polycrystalline samples with global $J_c$ of the order of $10^3$ A/cm$^2$ in self-field. To our knowledge, these values are about two orders of magnitude higher than those obtained on any other Fe(Se$_{0.5}$Te$_{0.5}$) phase in polycrystalline form so far. Remarkably, a quite flat behaviour versus the magnetic field is observed at T = 4.2 K, such that at $\mu_0 H$ = 7 T $J_c$ decreases only by a factor of about 2 with respect to the self-field value. The paper is organized as in the following way: after describing samples preparation (section 2) the detailed microstructural characterization by SEM analysis, Polarised Light Optical analysis and Electron back scattered diffraction analysis is given (section 3). Electrical (section 4) and magnetic characterization (section 5) are reported and global the critical current density evaluated by inductive measurements is compared with data obtained by direct transport measurements (section 6) and by magneto-optical analysis (section 7). All the results are compared with those obtained on a sintered sample prepared by a standard solid state reaction technique. Summary of the main results and conclusions are in the section 8.

**2. Experimental**

Fe(Se$_{0.5}$Te$_{0.5}$) powders were synthesized by the solid-state reaction method. High purity materials (Fe 3N+ powder, Se 5N powder and Te 5N lumps) were well mixed following the Fe(Se$_{0.5}$Te$_{0.5}$) stoichiometry in a glove box where the atmosphere was continuously maintained at H$_2$O/O$_2$ content lower than 1 ppm. Then the mixed powders were sealed in an evacuated Pyrex glass tube and slowly heated up to 450°C - 500°C for 100 hours to make sure that the reaction takes place. The result is a fine and homogeneous black powder. XRD analyses were performed on this first reaction product, showing the presence of a main PbO-type crystal structure, with small traces of compositional neighbouring compounds.

These powders were processed in order to obtain three different kinds of bulk samples, all being pellets of the order of magnitude of the centimetre.

1) Sintered sample (S): a pellet was made from the above described powders, closed in an evacuated Pyrex glass or quartz tube and heated at temperatures not higher than 800°C, temperature which is close to the melting point of the compound. The sintering time ranged from 20 to 40 hours, depending on the chosen temperature.

2) Melted sample (M): the powders were melted in a quartz crucible placed in a quartz tube under pure argon atmosphere at temperatures between 850° - 900°C followed by an air cooling. The obtained product was hard, compact and had a silvery lustre.

3) Melted +Annealed sample (M+A): the melted sample was then closed in an evacuated Pyrex glass or quartz tube and annealed in a furnace for 10 to 20 days at a temperature close to 550°C.

The relevant structural and superconducting parameters of these samples are listed in Table 1.

| Sample | Nominal composition | Synthesis process | Lattice constants | $T_c$ [K] | $\Delta T_c$ [K] | $\rho(300K)$ [m$\Omega$ cm] |
|---|---|---|---|---|---|---|
| S | FeSe$_{0.5}$Te$_{0.5}$ | Sintered | $a = 3.801$<br>$c = 6.043$ | 16.4 | 3.2 | 1.36 |
| M | FeSe$_{0.5}$Te$_{0.5}$ | Melted | $a = 3.802$<br>$c = 6.052$ | 16 | 3 | 0.63 |
| M+A | FeSe$_{0.5}$Te$_{0.5}$ | Melted + Annealed 550°C, 17 days | $a = 3.798$<br>$c = 5.964$ | 15.5 | 3 | 1 |

**Table 1**: Structural and superconducting parameters of 11 bulk samples

## 3. Microstructural characterization

*3.1 Scanning Electron Microscope Analyses*

Microstructure was analysed by a Scanning Electron Microscope (SEM: CAMBRIDGE S360), coupled with an electron dispersive spectrometer for chemical analysis (EDS: OXFORD Link Pentafet). Figure 1 shows the back-scattered SEM images of the three samples after metallographic preparation. A diffuse porosity distributed among the aggregated grains is observed in the S sample, whereas grain size ranges between ~ 10 μm and ~ 30 μm, comparable to the pore size. Interestingly the composition of the Fe(Se$_{1-x}$Te$_x$) phase is not homogeneous, as evidenced by the different grey scale intensity of the back-scattered images and confirmed by EDS analyses. Both lighter and darker grains exhibit homogenous compositions characterized by a [Se]/[Te] ratio that is on average ~ 0.63 and 1.17, respectively. This phenomenon is probably related to a thermodynamic instability of the Fe(Se$_{0.5}$Te$_{0.5}$) composition at 800°C, hence suggesting a miscibility gap in the pseudo-binary FeTe – FeSe system. The melted sample exhibits a very compact appearance, the porosity is reduced and the grain boundaries cannot be distinguished. The M sample is mainly constituted by a Fe(Se$_{1-x}$Te$_x$) matrix containing irregularly shaped grains of a secondary phase. Inside the matrix - lighter region in Figure 3.1, at centre - the [Se]/[Te] ratio is on average ~ 0.80. Fe(Se$_{1-x}$Te$_x$) is strongly enriched in Se around the secondary phases as evidenced by the g darker scale around the secondary phase grains. It is worth noticing that no interface can be detected moving along the different regions of the matrix. According to the SEM-EDS analysis, the secondary phase is probably constituted by Te-bearing Fe$_7$Se$_8$. This phase is largely dissolved during the annealing treatment, as observed in the back-scattered images of the M+A sample (Figure 1, on the right); in this sample the Fe(Se$_{1-x}$Te$_x$) constitutes lighter Te-enriched regions ([Se]/[Te] ~ 0.87) or darker Se-enriched ones ([Se]/[Te] ~ 1.36), but no interface separates the different zones. This demixing is similar to what observed in the S sample, indicating that the thermodynamic instability in the FeTe – FeSe system extends down to 550°C. Light oriented precipitates of a secondary phase are present within the matrix; SEM-EDS analysis reveals that this phase is strongly Fe-depleted and the [Se]/[Te] ratio is 1:3. Attempts to identify this phase were unsuccessful: this is probably a new ternary phase belonging to the Fe-Te-Se system and its chemical formula can be referred to as Fe$_2$Ch$_3$ (*Ch*: chalcogen).

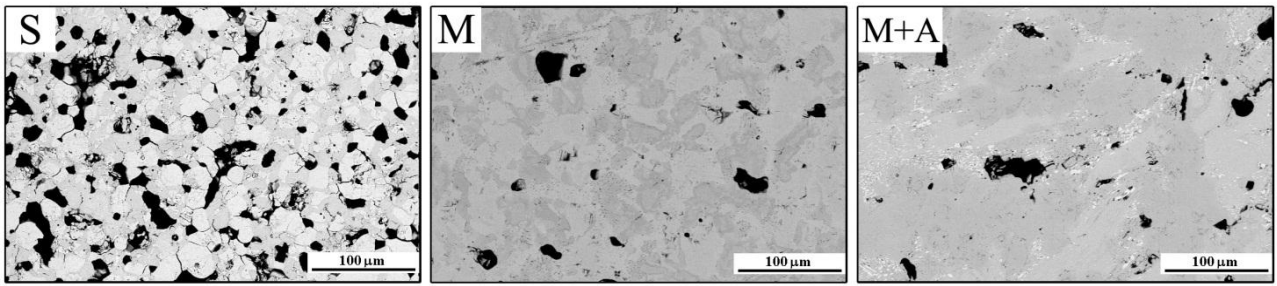

**Figure 1**: Back-scattered SEM images of the three samples. Left: S sample; middle: M sample; right: M+A sample. Black regions are pores within the samples.

*3.2 Polarized Light Optical Analysis*

In order to determine the grain size and shape, the samples were observed using a polarized light optical microscope (PLOM). Grain size measured on the S sample with this technique is in good agreement with that obtained by SEM images, confirming the reliability of the technique. In the M sample linear grain size reaches macroscopic values, ranging from ~ 100 μm up to ~ 1 mm, in some cases; grains with different orientations can be clearly distinguished, since they are characterized by very different interference colours (Figure 2). The annealing treatment induces a re-distribution of the chemical elements as well as a re-crystallization of the sample; as a consequence the linear grain size in the M+A sample still ranges from ~ 100 μm up to ~ 1 mm, but on average it is slightly decreased.

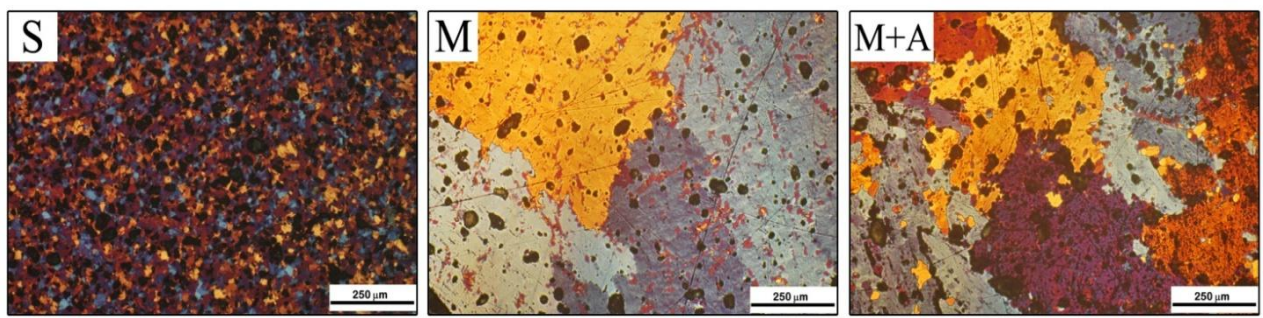

**Figure 2**: PLOM images of the three samples. Left: S sample; middle: M sample; right: M+A sample. Darker regions are pores within the samples.

The microstructures of the M and M+A specimens strongly differ from those observed in samples belonging to the 1111-family, where 1111 crystals are embedded within the non-superconducting wetting FeAs phase [9,10]. In fact no evidence for precipitation at grain boundary can be detected in both M and M+A specimens, even at higher magnification (see Figure 3) and by EBSD analysis (see § 3.3 below), and the different $Fe(Se_{1-x}Te_x)$ grains are closely connected. This different behavior can be ascribed to the different preparation technique and thermodynamic of the 11 and 1111 compounds. In fact an optimal distribution of the reacting elements is achieved by melting during the synthesis of 11 type compounds and the formation of competing phases, induced by local compositional inhomogeneity, is hindered. Conversely the synthesis of 1111-type compounds is typically carried out by solid state techniques, that can be affected by not-homogeneous distribution of the reactants whereas the formation of the desired phase is controlled by diffusion. In addition the formation of the quaternary 1111 compound at high temperature must compete with that of simpler phases, such as the low-melting FeAs binary compound, found as a wetting phase separating the 1111-type crystals [9, 10]. This kind of competition is suppressed during the synthesis of $Fe(Se_{1-x}Te_x)$, where all the forming phases are fundamentally binary compounds. These secondary binary compounds can be easily detected as dispersed intra-grain precipitates (Figures 1 and 2), but no evidence for their occurrence can be detected at grain boundaries.

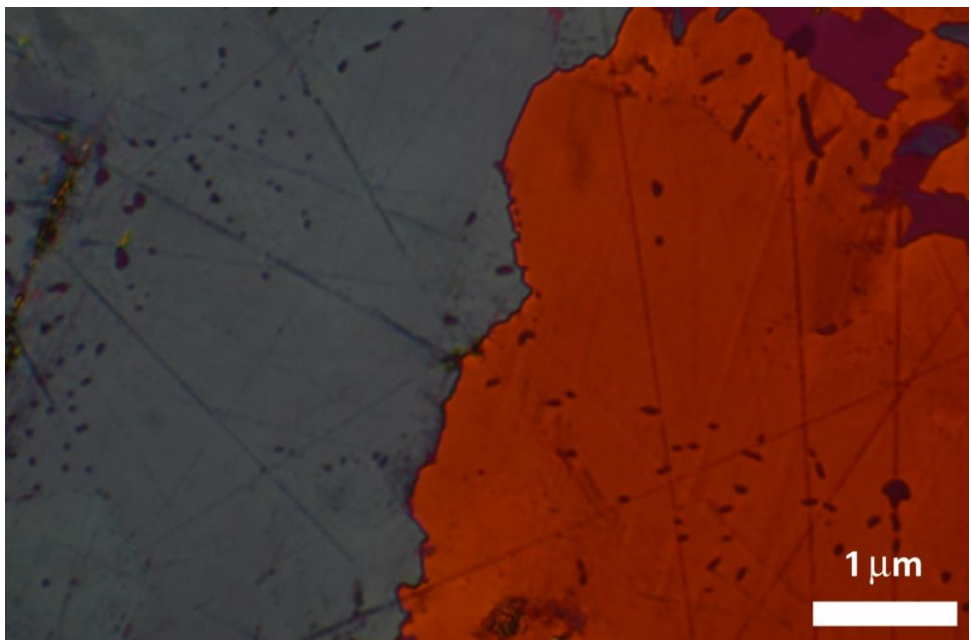

**Figure 3**: High magnification PLOM image collected on the M sample and showing the region separating two different $Fe(Se_{1-x}Te_x)$ grains: no evidence for precipitation at grain boundary can be detected.

*3.3 Electron Back Scattering Diffraction Analysis*

In order to investigate possible crystallographic relationships among neighbouring grains, electron back scattered diffraction (EBSD) measurements were carried out on the M+A sample. The EBSD investigation [23] was carried out using an Inca Crystal 300 EBSD system added to a SEM LEO (model EVO 50) with $LaB_6$ gun. Through this technique the electron beam scans a selected area of the sample; the resulting Kikuchi electron diffraction patterns [24] are indexed automatically to produce a crystallographic orientation map. The different orientations are then highlighted according to the colour-coded stereographic triangle for the studied phase. To perform this experiment, the crystallographic cell of the $Fe(Se_{0.5}Te_{0.5})$ compound was entered the software database for the phase identification.

First, the EBSD analysis was performed on a cleaved surface of the M+A sample. High quality diffraction patterns were acquired on a flat area (400×500 μm$^2$; beam step ∼ 2 μm$^2$) of the sample. A map based on the image quality (IQ) data (Figure 4(a)) reveals a contrast related to local variations of surface quality, crystallinity and orientation (dark and light regions). Lighter regions are observed which highlight homogeneous zones, shaped as grains, in the greyish to dark matrix. Figures 4 (b)-(d) show the EBSD crystallographic orientation maps (COMs) along [0 0 1], [1 0 0] and [1 1 0] crystallographic directions

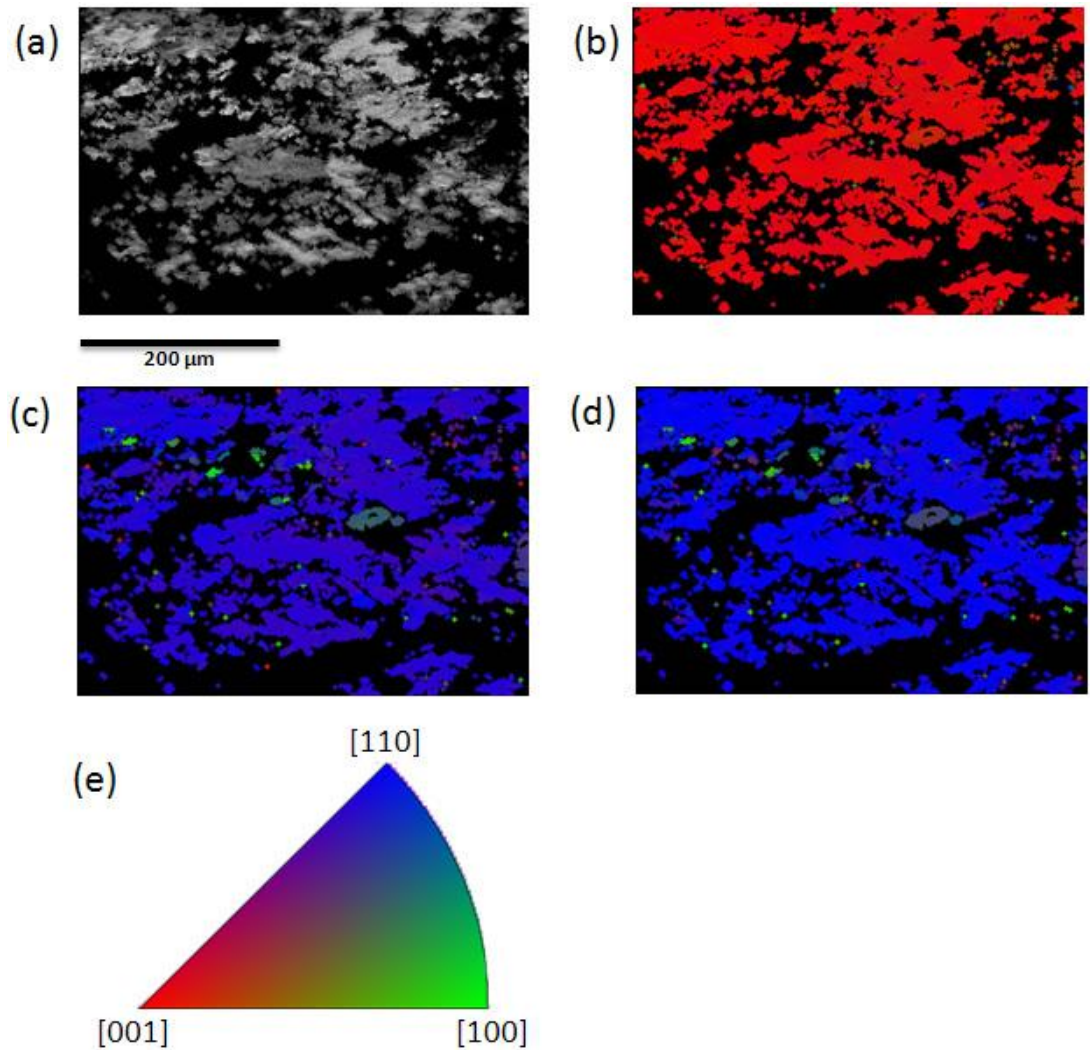

**Figure 4:** EBSD crystal orientation mapping of the Fe(Se$_{0.5}$Te$_{0.5}$) M+A cleaved sample collected at 20 kV accelerating voltage. The map step size is ∼1.5 μm. (a) pattern quality map, (b)–(d) COMs in [0 0 1], [1 0 0] and [1 1 0] directions, respectively. The crystallographic orientations are given in the colour-coded stereographic triangle for Fe(Se$_{0.5}$Te$_{0.5}$) phase (e).

In agreement with light polarized optical observations these results prove that the M+A sample is constituted of large grains of a few hundreds microns characterized by good crystallinity.

Similar analyses were carried out on samples metallographically prepared (diamond paste down to 1 μm, colloidal silica and final chemical etching). A good diffraction pattern was mainly observed on domains with *c*-axis parallel to the normal to the sample surface, the same as observed on cleaved sample, suggesting that the plane orthogonal to the *c*-axis tends to be more easily polished giving good Kikuchi pattern. These regions have areas of less than half mm$^2$ confirming the presence of grains of few hundred of microns, as observed by PLOM images.

Figure 5 shows the EBSD crystal orientation mapping for a scanned area of 370×280 μm$^2$ which exhibits a good pattern quality image. Figures 5 (a)-(c) present COMs in [0 0 1], [1 0 0] and [1 1 0] crystallographic directions. These measurements confirm that the scanned area is mainly *c*-axis oriented. However, also regions with different colours appear indicating that grains with different orientations are present. In order to evaluate the misorientation between the main and the neighbouring grains the EBSD data points were analyzed. A grain map was defined considering as part of the same grain neighbouring pixels with a misorientation less than or equal to 5°, in addition a grain consisting of less than five pixels was excluded from the grain size calculation. The resulting grain map, in Figure 5 (d), shows a main grain surrounded by smaller grains. Each grain is assigned a discrete colour to differentiate it from the neighbouring grains.

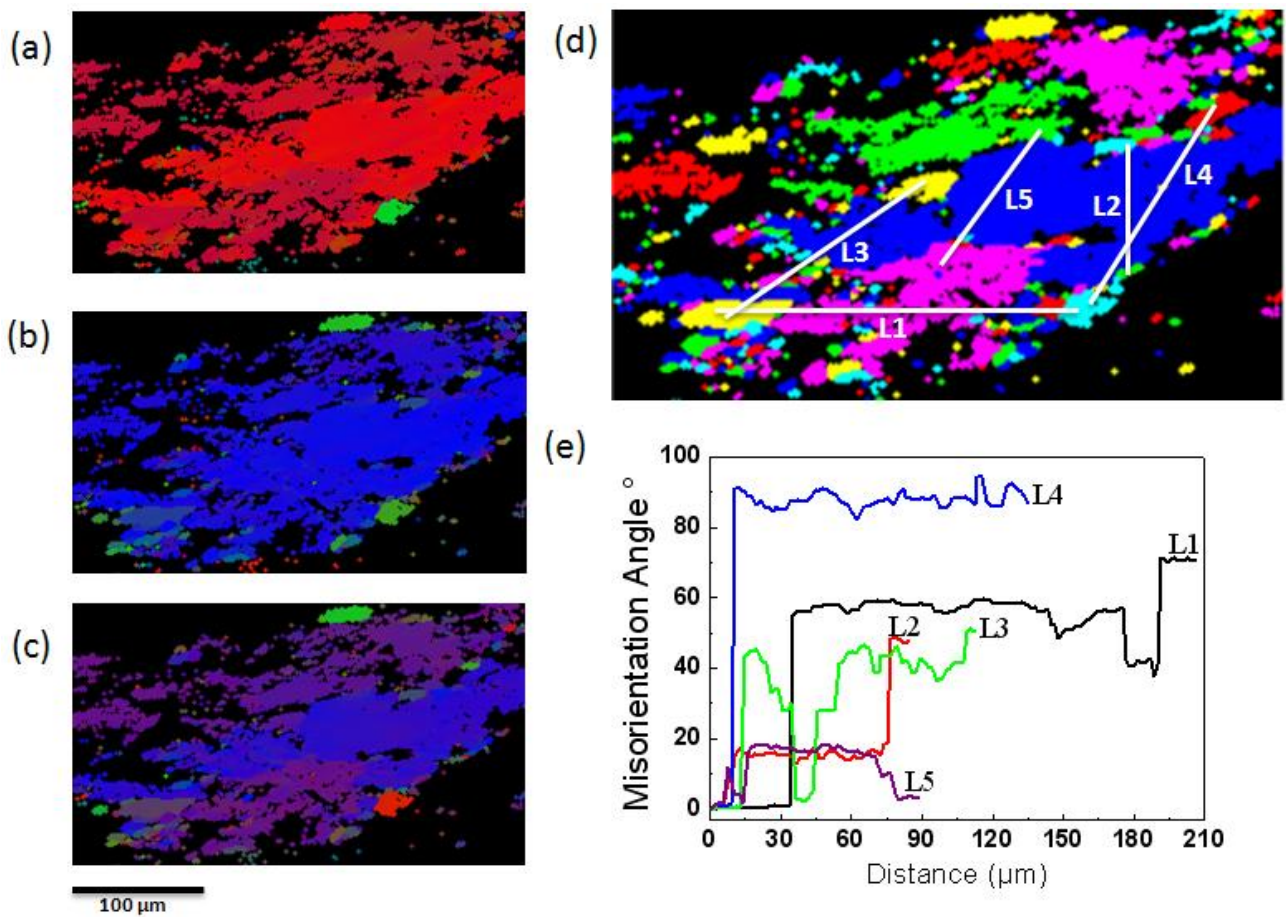

**Figure 5:** (a)–(c) EBSD crystal orientation mapping of polished the M+A sample collected at 20 kV accelerating voltage in [0 0 1], [1 0 0] and [1 1 0] directions, respectively, (d) grain map calculated with a misorientation angle threshold of 5° and, (e) variations of misorientation angles for profile scans indicated in the grain map (d).

The misorientations from one grain to another can be quantified [25] by line scans of misorientation. The plots of line-scans from Figure 5 (d) show grains separated by high (more than 15°) boundary angles (Figure 5 (e)). This analysis suggests that in these samples grains with large misorientation grow strongly interconnected without exhibiting spurious phases at the interfaces, as also confirmed by compositional SEM analysis.

## 4. Electrical characterization

Figure 6 shows the resistivity as a function of temperature of the S , M and M+A samples.

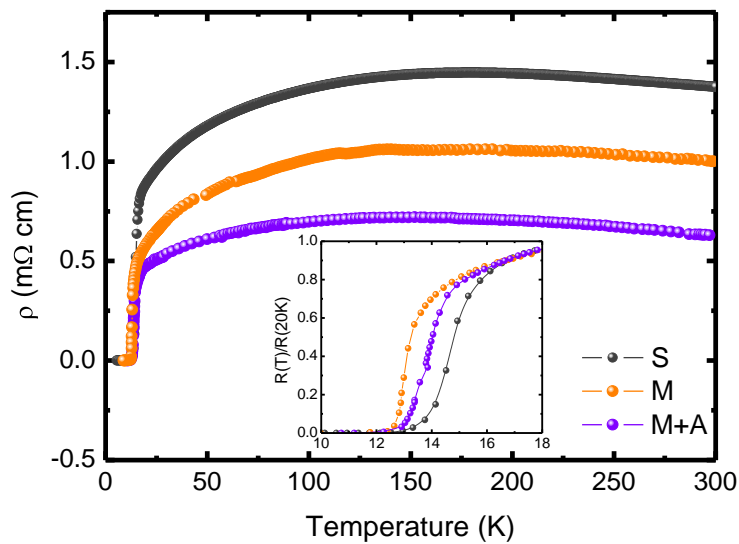

**Figure 6:** Resistivity as a function of temperature of the samples S, M and M+A.

The temperature dependence does not vary significantly from sample to sample, but the resistivity values in the melted samples are significantly lower with respect to the sintered sample consistent with the increase of the density discussed in the previous section. On the other hand, the M+A sample has larger resistivity than the M sample: this feature has been systematically observed and could be due to the different composition of the matrix of the two samples: indeed the M sample presents regions more Se enriched in respect to M+A sample and the resistivity of the 11 compounds progressively decreases by increasing the Se content [26].

The inset shows the magnification of the region around the superconducting transition. We can see that the $T_c$ onset evaluated at 90% of the normal state resistivity is the largest for the S sample (16.4 K) and then it progressively decreases to 16.0 K and 15.5 K for the M and for the M+A samples, respectively. Suppression of superconductivity in Fe chalcogenides by annealing has been recently investigated by neutron diffraction analysis [27]. After annealing the chalcogen ion's z parameter has been observed to change in the opposite direction as that produced by pressure, which suggests that annealing relaxes the stress in the structure inducing a suppression of $T_c$. However, looking at Figure 6, the amplitude of the transition becomes sharper and sharper going from the S to the M to the M+A so that all the samples exhibit zero resistivity nearly at the same temperature which is around 12.5 - 13 K.

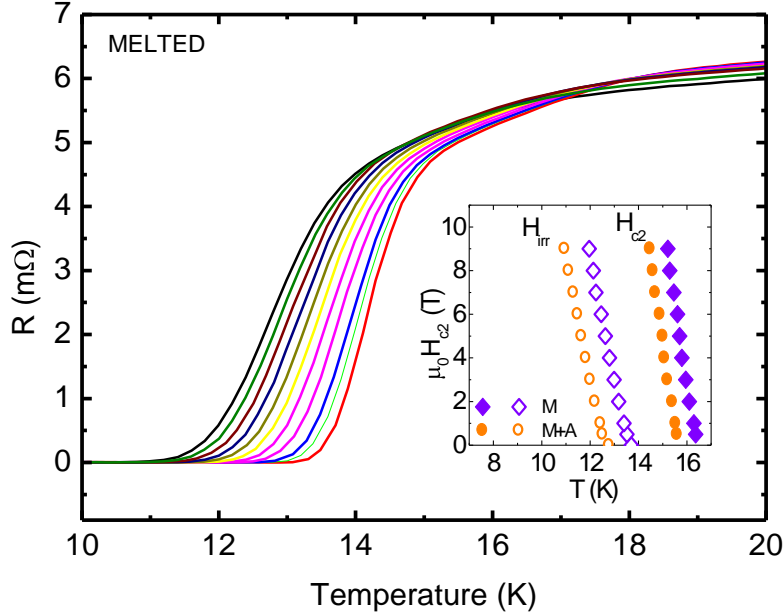

**Figure 7:** Magnetoresistivity as a function of temperature of the M sample in applied magnetic field up to 9 T. Inset: the upper critical field, $\mu_0 H_{c2}$, and the irreversibility field, $\mu_0 H_{irr}$, for the M and M+A samples.

Magnetoresistance measurements in applied magnetic field from 0 to 9 T of the M sample are plotted in Figure 7. The transition progressively shifts to lower temperature with a moderate broadening with increasing the field. A very similar behaviour is observed in the M+A sample. In the inset of Figure 7 the upper critical fields $\mu_0 H_{c2}$, evaluated at 90% of the normal state resistivity, and the irreversibility fields $\mu_0 H_{irr}$, evaluated at 10% of the normal state resistivity, are plotted for the M and M+A samples. $\mu_0 H_{c2}$ curves steeply increase with decreasing temperature and a slope $-\mu_0 dH_{c2}/dT$ of about 7.6 T/K can be evaluated for both the samples. This value is larger than those reported in polycrystalline materials (6 T/K)[22] but much lower than the huge values reported in single crystals (30 T/K)[1] and thin films (500 T/K) [28]. Indeed, inhomogeneity and grain misalignment smear the transition and mask the huge $\mu_0 H_{c2}$ of this compound. Also $\mu_0 H_{irr}$ increases sharply with decreasing the temperature, with nearly the same slope as $H_{c2}$, indicating that the transition broadening is negligible in these samples.

## 5. Magnetic characterization

Magnetization measurements were performed in a dc-superconducting quantum interference device SQUID magnetometer (Magnetic Properties Measurement System by Quantum Design) on slab-shaped samples (typically ~ 2 mm long and thickness $d$ about few hundred microns) with the magnetic field applied in the slab plane. In Figure 8 we plot the temperature dependence of the zero-field-cooled (ZFC) dc susceptibility measured in an applied field $\mu_0 H = 1$ mT. All the samples show nearly the same onset of the superconducting transition ($T_{c\text{-}on\text{-}mag}$, indicated by a red arrow) around 12.5 K which corresponds to the temperature of vanishing resistivity (see the inset of Figure 6).

As the temperature decreases, the differences among the samples become significant. In the case of S sample the susceptibility $\chi$ starts to decrease slowly and full shielding is not reached even at the lowest temperature. This behaviour suggests the presence of both inhomogeneity and weak inter-grain connections. The M sample shows an enhanced diamagnetic behaviour even if the superconducting transition remains quite broad. On the contrary; the M+A sample exhibits a much sharper magnetic transition and the full shielding is reached below 9 K. Such behaviour has been observed only in the best quality single crystals to the best of our knowledge [29,30,31,32,33]. Bigger and better interconnected grains are at the origin of these strongly improved superconducting properties. Summarizing, the M and M+A samples show enhanced diamagnetic response which suggests that the melt process, especially with subsequent annealing, improves the homogeneity and the connectivity between grains thus optimizing the superconducting properties.

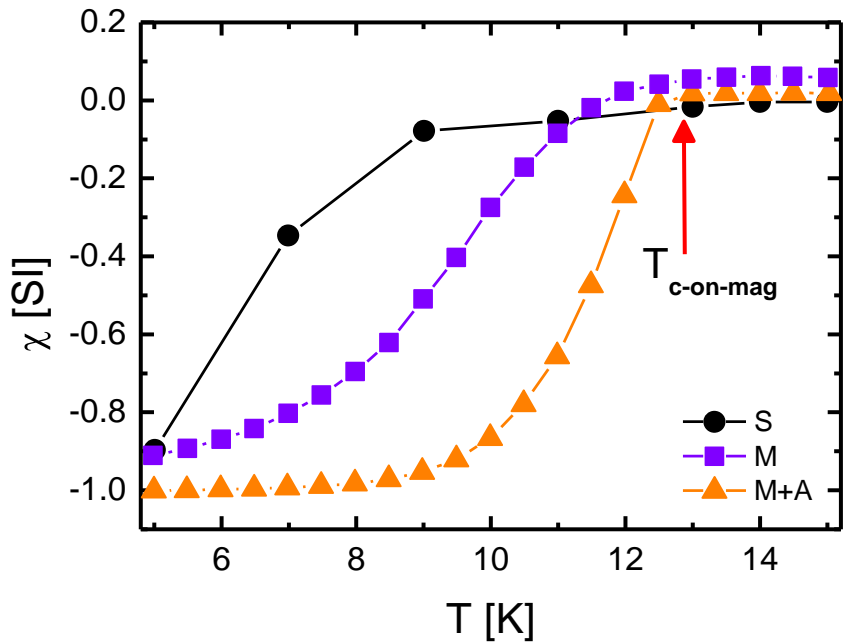

**Figure 8**: Temperature dependences of ZFC susceptibilities for the S, M and M+A samples measured at $\mu_0 H = 1$ mT. The red arrow indicates the onset temperature of the magnetic superconducting transition.

In Figure 9 the hysteresis loops measured at T = 5K, in fields up to $\mu_0 H = 5$T are shown. The intrinsic signal is superimposed to a spurious one related to ferromagnetic impurities. It is worth noticing that: *i*) the weight of this spurious ferromagnetic component is strongly reduced in the M+A sample, and *ii*) the amplitude of the hysteresis loops greatly increases while passing from S to M and from M to M+A samples.

In a granular superconductor, the magnetization depends on the sum of inter and intra-grain contributions. To separate these two contributions, remnant magnetization measurements, $m_R$, are particularly useful

[34,35,36]. $m_R$ as a function of the maximum applied magnetic field $H_m$ is shown in Figure 10. It is important to outline several interesting features. First of all, in the S sample $m_R$ increases very quickly with magnetic field indicating a nearly zero effective first critical field (see inset in Figure 10). We recall that in a type II superconductor, the first critical field is the field at which $m_R \neq 0$ due to the first vortex penetration. Instead, in a granular superconductor we have to consider the effective first critical field that corresponds to the applied field value at which the first vortex penetrates the sample weak links and for this reason it is straightforwardly called first Josephson critical field ($\mu_0 H_{c1J}$). This value is larger in the M sample ($\mu_0 H_{c1J} \sim 1$ mT) and it is maximum in the M+A sample ($\mu_0 H_{c1J} \sim 4$ mT) as shown in the inset of Figure 10. It is worth noticing that for a Fe(Se$_{0.5}$Te$_{0.5}$) single crystals $\mu_0 H_{c1}$ values of 4.5 mT and 2 mT in the *c*-axis and *ab*-plane direction, respectively, have been reported [37], very close to the value extracted for the M+A sample, suggesting the presence of "strong" links between grains. $m_R$ ($H_m \to \infty$) is maximum for the M+A sample. Because the grain size is not increased by the annealing (see section 3), larger asymptotic $m_R$ values suggest the improvement of either inter-grain or intra-grain critical currents or both [34].

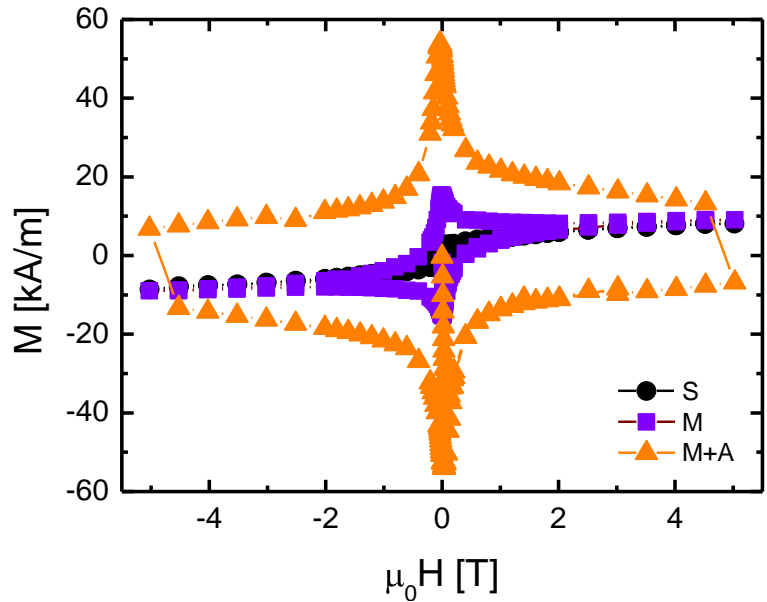

**Figure 9**: Hysteresis loops measured at T = 5 K on the S, M, M+A samples.

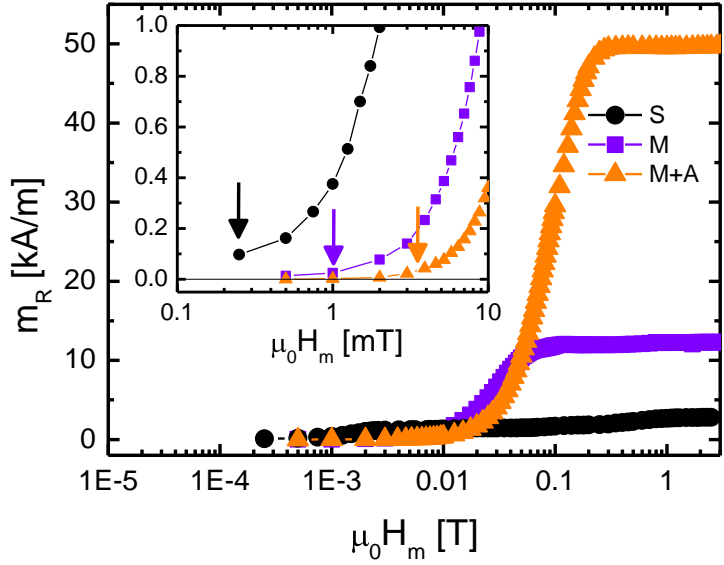

**Figure 10:** Remnant magnetization as a function of the maximum applied field $H_m$. In the inset a magnification of the low field region is shown. The arrows indicate the effective first critical field.

In order to evaluate the global critical current from these magnetic measurements, we examined the derivative of the remnant magnetization $dm_R/dH_m$ which is presented in Figure 11 for all the samples. We recall that in fully connected samples $dm_R/dH_m$ has the shape of the derivative of a sigmoidal function. In granular samples, separated peaks and/or several shoulders and/or kinks corresponding to the presence of different superposed peaks appear, as in our case. The presence of several peaks suggests a hierarchy of distinct different paths of current flow that are successively broken by the magnetic field. In a basic approach, the global critical current is derived from the Bean critical field ($H^*$) at which the sample is fully penetrated by vortex: $J_c^{global} = 2H^*/d$ [38] where $d$ is the sample thickness. This field is the position of the left-hand first peak of the $dm_R/dH_m$ curve along the $H_m$ axis [34,35].

As it may be observed in Figure 11, while in the S sample $H^*$ is well identified by the peak at low field, in the M and M+A samples the peaks related to the global current shift to higher field and overlap with peaks related to current flowing on smaller scale. Therefore, we safely identified $H^*$ by the shoulder indicated by the arrows in the Figure 11. For the M+A sample we fitted $dm_R/dH_m$ by the sum of three Gaussian functions. The choice of three Gaussian curves does not have a particular physical meaning: it simply represents a situation in which connected regions (more than two) whose dimension depend on the applied magnetic field are present. The so obtained $J_c^{global}$ values result to be about 0.2, 5.0 and 5.5 kA/cm$^2$ for the S, M, M+A respectively. The values obtained for the M and M+A samples are of the same order of magnitude as the values measured by transport measurements (*vide infra*), and are the highest obtained in bulk samples of 11-phase so far to the best of our knowledge.

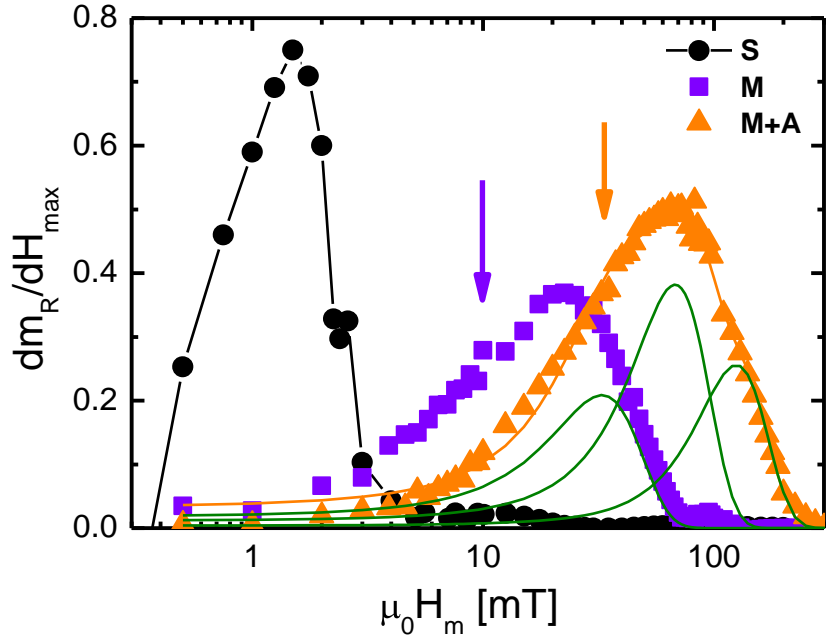

**Figure 11:** d$m_R$/d$H_{max}$ for the three studied samples. The continuous lines are numerical fits obtained by using the sum of three Gaussian functions. The green lines represents each component of the full fitting function for the M+A sample as example.

## 6. I-V characterization

Transport $J_c$ measurements in polycrystalline samples are made difficult by the high values of current needed, with the related heating and instability problems which can arise in relatively short, bulk samples without any stabilization. Therefore great care was put into the four-probe transport $I_c$ measurements. The measurements were performed at 4.2 K in a 7 T split-coil superconducting magnet on the M+A sample. In order to make the sample more robust to mechanical strains, we embedded it into a two-component epoxy in parallelepiped shape. Then, we polished the exposed surface to make the sample thinner, reaching an average cross-section of 0.02 cm$^2$, that allowed the use of relatively low currents (below 100 A). In order to avoid any local damage on the sample, for example by soldering, the current contacts were realized mechanically by pressing the current leads – made with two Cu sheets - on the sample itself using two thick insulated plates acting like a 'sandwich'. To lower further the contact resistance, a thin layer of In was placed between the sample and the current leads thus achieving a very good electrical contact ($\approx 10^{-6}$ $\Omega$cm$^2$). In Figure 12 (a) we show the picture taken with a polarized light microscope – as described in section 3 – on the relevant portion of sample before measuring it by transport. A sketch of the voltage taps and of the position of the current leads is also shown, to better clarify that during this global measurement the current was flowing through portions of the sample presenting different orientations, therefore crossing several grain boundaries.

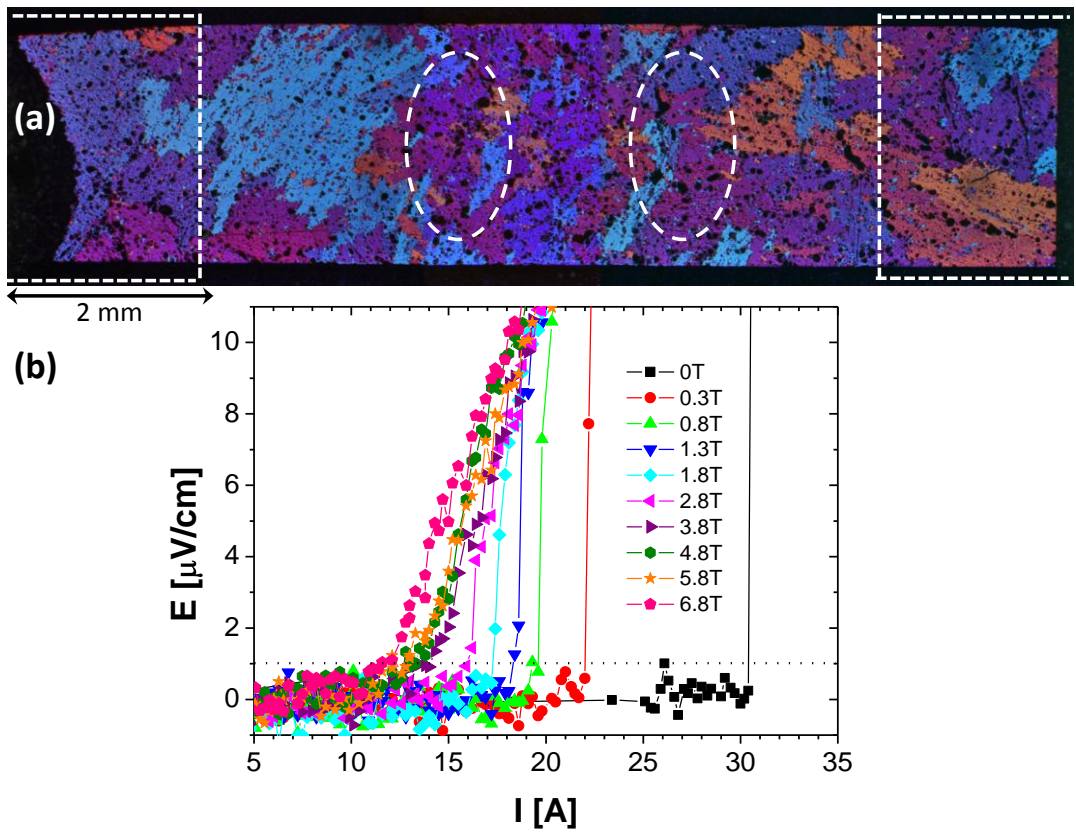

**Figure 12:** (a): polarized light image of the M+A sample prepared for transport measurements: the position of the voltage and current pads are sketched. (b): I–V characteristics of the M+A sample for some selected magnetic fields.

In Figure 12 (b) some selected *I-V* transitions are reported: for lower fields, say below 1 T, we observe sharp transitions which can be ascribed either to an overheating occurring at the interface between the current leads and the superconducting phase, or to a local instability. This is a very common problem arising during such measurements especially when the samples are bulks without any stabilization: it is possible that for the data below 1 T we slightly underestimate the actual critical current. The transitions above 1 T are not sharp and show the correct curvature, making it possible to extract a reliable value of critical current.

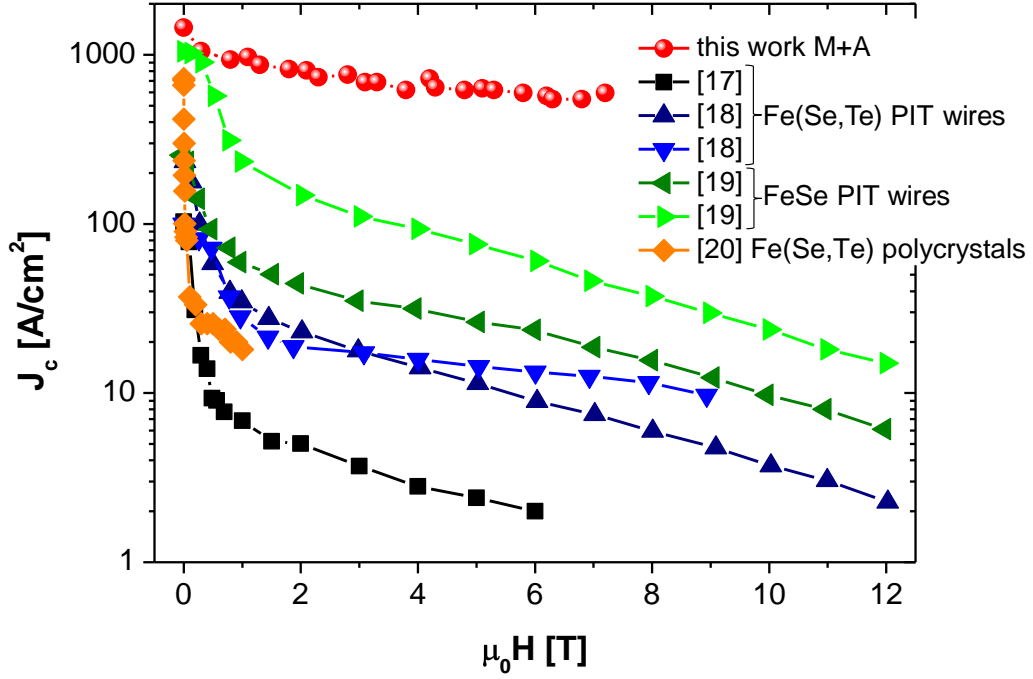

**Figure 13:** Transport $J_c$ of the M+A sample compared with data reported for bulk 11 samples.

In Figure 13 we report the transport $J_c$ as a function of the magnetic field at 4.2 K extracted from the *I-V* curves as defined by the usual 1 µV/cm criterion. Together with our data we report a comparison with recent data measured on different massive samples of the phase (11), either bulk samples [22] or PIT mono- or multi-core wires [18, 19, 20]. We point out that the transport $J_c$ values of the M+A sample are the highest ever reported for this phase (11). In particular, the most striking and important feature of our data is the magnetic field dependence which is extremely weak, going from 1400 to 600 A/cm$^2$ as the field increases from 0 to 7 T. This is a clear indication of the fact that the connection between the grains is strongly improved through our preparation technique. Despite the data below 1 T might be underestimated as already discussed, still the curve remains extremely flat between 1 and 7 T where the transitions are absolutely reliable.

In addition, by comparing magnetic and transport measurements, we can try to separate the intergranular and intragranular contribution to the magnetization in order to extract the intragranular $J_c$ of our sample. In fact, we can use transport data to calculate the magnetization signal which corresponds to the intergranular (global) component, and then subtract it from the total magnetic signal to obtain the intragranular contribution only. By doing this analysis we evaluate that the contribution of the global current to the hysteresis loop never exceeds the 20% of the total signal. If we hypothesize a mean grain size of the order of 150 µm as evaluated by the polarized light images of the M+A sample, intragranular $J_c$ can be evaluated. In Figure 14 the intragranular $J_c$ versus magnetic field as obtained in such a way is shown. $J_c$ is of the order of $10^5$ A/cm$^2$ at zero field and decreases down to $2\times10^4$ A/cm$^2$ at 4 T. These values are comparable with those reported in single crystals [33, 39, 40] as shown in Figure 14. However, they are much lower than those reported in the 11 thin films [41], for which $J_c$ remains well above $10^5$ A/cm$^2$ at high field, suggesting that pinning mechanisms in this phase can be strongly improved.

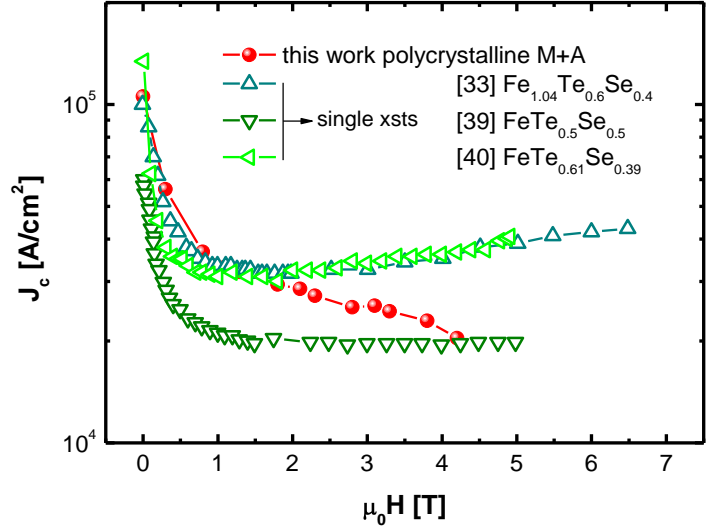

**Figure 14:** Intragrain $J_c$ evaluated by magnetization measurements for the M+A sample as explained in the text. For comparison data on Fe(Se,Te) single crystals are reported.

## 7. Magneto-optical characterization

Magneto-optical technique on the base of garnet film indicators with in-plane magnetization [42, 43, 44] was used for visualization of flux behaviour in the three polycrystalline S, M and M+A samples. All samples were cut in rectangular prisms with size (2-3) x (3-4) mm and thickness about 0.3 mm. The widest surface of each sample was polished to shiny condition and investigated in a polarized light microscope. Magneto-optical indicator, which picks-up a map of flux distribution and transfer it to a visible spectrum, was placed on wide polished surface of sample. Polycrystalline samples were installed in a continuous flow portable cryostat with an optical window and cooled down to temperature below $T_c$ transition. External field was applied from electromagnetic coil perpendicular to wide surface of samples. The S sample did not show acceptable MO contrast, while the M and M+A samples showed good contrast in both ZFC and FC regimes. Figures 15 and 16 demonstrate polarized optical and MO images taken on the M and M+A samples. As shown from polarized optical image, the melted sample M has large grains with different orientation. As visible from MO, the intergrain areas are places where magnetic flux penetrates with increasing magnetic field in ZFC regime. Such obstacles limit the global bulk current in magnetic and transport measurements. The estimation from MO image gives us the value of global current $J_c$ of the order of 1-2x10$^3$ A/cm$^2$ at T = 6 K and well correlated with transport measurements. However, there are regions in which the local current is higher. It may be due to large and well connected grains which appear in bright contrast in FC regime. Additional annealing procedure makes structure of the M+A sample look different, with smaller and better connected grains. MO image shows remarkable "roof pattern" on the left side of sample and testify bulk current in this area. MO contrast is very inhomogeneous over the sample and indicates that the current distribution is also non-uniform. The right side of sample has weaker critical current than the left side and only part of the cross section can carry transport current. This indicates that there is a lot of room for an improvement of sample structure and, as a result, for increasing the critical current.

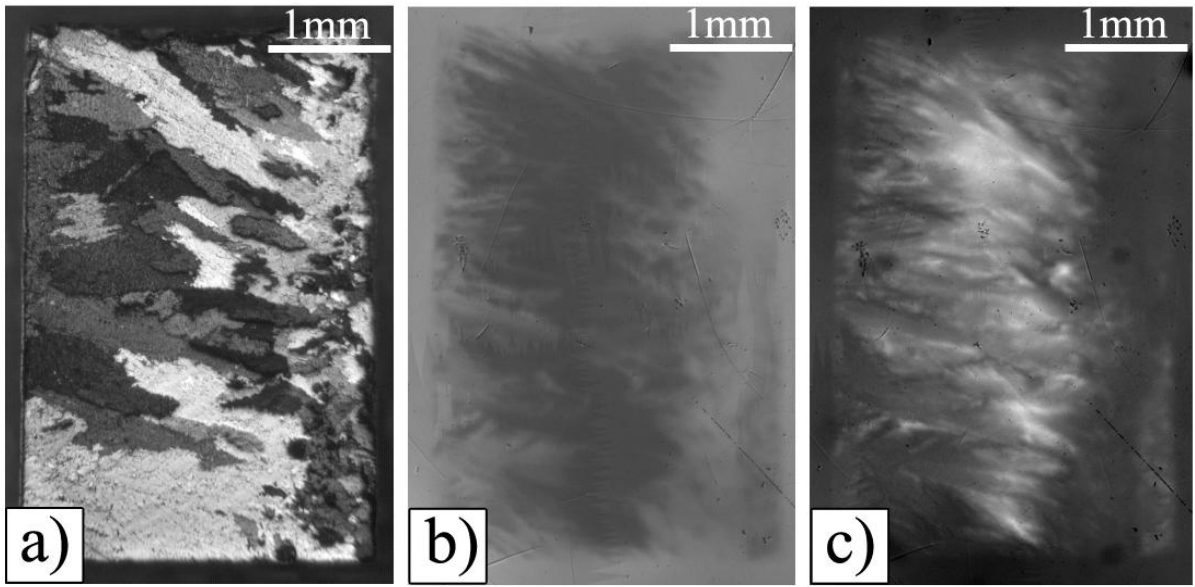

**Figure 14:** Polarized optical image of the melted (M) sample and Magneto-optical images taken at T = 6 K: b) in ZFC and field 9 mT and c) in FC in zero field after cooled down in field 120 mT.

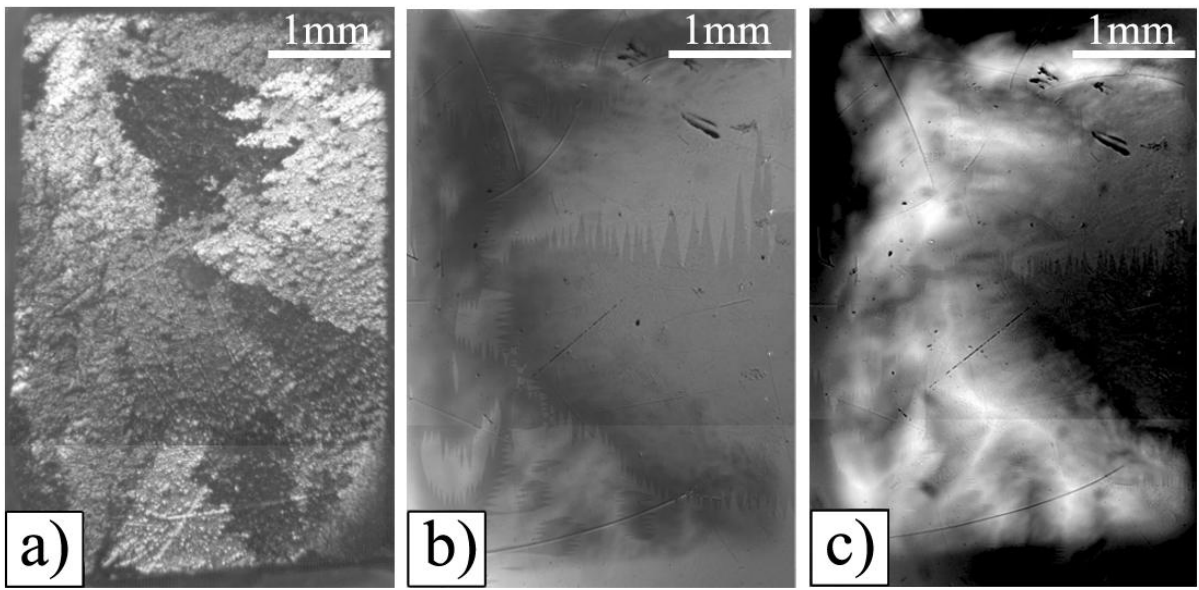

**Figure 15:** Polarized optical image of the melted + annealed sample (M +A) and Magneto-optical images taken at T = 6 K: b) in ZFC and field 9 mT and c) in FC in zero field after cooled down in field 120 mT

## 8. Conclusions

A new multi-step method was developed to synthesize Fe($Se_{0.5}Te_{0.5}$) bulk samples with improved superconducting properties, involving a melting process and a subsequent annealing treatment.

The samples obtained by melting are much more compact and denser in comparison with the standard sintered samples, with large and strongly connected Fe($Se_{1-x}Te_x$) grains; the subsequent annealing treatment increases the sample homogeneity without an appreciable increase of the grain size. In both cases no evidence for precipitation at grain boundaries can be detected by structural and microstructural analysis, whereas large misorientation characterizes adjacent grains .

The M+A samples present optimal $T_c$ and exhibit sharp magnetic transitions and full shielding. The resistive transition in applied field does not broaden significantly and the $\mu_0 H_{c2}$ and $\mu_0 H_{irr}$ slopes are among the highest measured in polycrystalline materials. The hysteresis loops strongly enlarge after annealing which implies a significant improvement of the critical current, both inter and intra-grain.

By the remnant magnetization of the M and M+A samples a global current larger than $10^3$ A/cm$^2$ is estimated but this analysis also suggests that the current distribution is not uniform and there exist different current paths that are progressively interrupted by an increasing magnetic field. These results are confirmed by magneto optical imaging that shows patterns are very inhomogeneous over the samples suggesting that grain boundaries still act as severe block .

Finally, transport measurements of the M+A sample provide a value of critical current in self field of $1.4 \times 10^3$ A/cm$^2$ in substantial agreement with remnant magnetization and MO analyses. The most important result is that the transport $J_c$ value is only weakly affected by the applied magnetic field and a value of 600 A/cm$^2$ (12 A) was evaluated at 7 T. This value is more than one order of magnitude larger than the ones previously obtained in the 11 phase (see Figure 13), but it is still one order of magnitude lower than that reported for K doped 122 tapes $3.5 \times 10^3$ A/cm$^2$ in 10 T [15] and wires $10^4$ A/cm$^2$ in 10 T [17].

The reduced $J_c$ values in respect to the 122 phase may arise mainly by intrinsic reasons, among them the lower $T_c$, as proved by the fact that single crystals of the 11 phase exhibit the lowest $J_c$ in comparison with iron-pnictide superconductors (1111 and 122 phases).

On the other hand the precipitation at grain boundary, which causes current blocking, is prevented by the melting synthesis technique in 11-type compounds. Nevertheless compositional inhomogeneity, driven by thermodynamic instabilities, plays a major on the superconducting properties [45] and can originate local inhomogeneous $J_c$.

However in the last year a step-like $J_c$ improvements have been reached in the 122, 1111 and, with the present work, in the 11 bulk samples, indicating that there is enough potential for iron based superconductor applications, thus motivating further research and developments of these materials.

**Acknowledgements**

This work has been supported by FP7 European project SUPER-IRON (grant agreement No.283204). R.F. acknowledges support by the MAMA European project, FP7/2007-2013 under Grant agreement No. 264098. We are grateful to David Larbalestier for valuable remarks and discussion.